# Synthesis and characterizations of arsenic doped FeSe bulks

Priya Singh[1], Manasa Manasa[1], Mohammad Azam[1], Tatiana Zajarniuk[2], Konrad Kwatek[3], Tomasz Cetner[1], Andrzej Morawski[1], Jan Mizeracki[1], Shiv J. Singh[1*]

[1]*Institute of High Pressure Physics (IHPP), Polish Academy of Sciences, Sokołowska 29/37, 01-142 Warsaw, Poland*

[2]*Institute of Physics, Polish Academy of Sciences, Aleja Lotnikow 32/46, 02-668 Warsaw, Poland*

[3]*Faculty of Physics, Warsaw University of Technology, Koszykowa 75, 00-662 Warsaw, Poland*

\*Corresponding author:

Email: sjs@unipress.waw.pl

## Abstract

FeSe(11) family has a simple crystal structure belonging to iron-based superconductors (FBS) and has many stable phases including hexagonal and tetragonal structures, but only the tetragonal phase exhibits the superconductivity. In this study, we have investigated the effects of chemical pressure induced by As-doping at Se-sites in the FeSe system by preparing a series of FeSe$_{1-x}$As$_x$ ($x$ = 0.005, 0.01, 0.02, 0.05, 0.1, and 0.2) bulks. A broad characterization has been performed on these samples using structural, microstructural, transport, and magnetic measurements. The obtained lattice parameters are increased by As-doping, which suggests the successful insertion of As at Se-sites into the tetragonal lattice for low doping contents up to 5%, whereas the higher As-substitution appears in the form of the FeAs impurity phase. The temperature dependence of the resistivity of all samples has similar behaviour and depicts the highest onset transition temperature of around 11.5 K, but the zero resistivity is not reached until the measured temperature of 7 K, which could be due to the presence of the impurity phases. Our study suggests that a dopant with a large ionic radius, i.e., Arsenic, promotes the formation of the hexagonal phase of the 11 family and is effective for a small amount of doping level for the superconducting properties, whereas higher As-doping levels reduce the superconducting properties.

# 1. Introduction

The iron-based superconductor (FBS) was discovered in 2008 through F-doped LaOFeAs, exhibiting a superconducting transition temperature $T_c$ = 26 K [1]. Following this report, many compounds have been discovered under this high-$T_c$ material and the transition temperature ($T_c$) has been enhanced up to 58 K for F-doped SmFeAsO [2], [3]. In addition, the $T_c$ of F-doped LaOFeAs is increased up to 43 K under the applied external pressure [4]. These high-$T_c$ superconductors can be classified into 6–7 families following their crystal structure [5], [6], [7], [8] : $RE$OFeAs (1111) ($RE$ = rare earth), $A$Fe$_2$As$_2$ (122) ($A$ = Ba, K, Ca) (122), CaKFe$_4$As$_4$ (1144), and LiFeAs (111), thick perovskite-type oxide blocking layers, including Sr$_4$V$_2$O$_6$Fe$_2$As$_2$ (42622), Sr$_4$Sc$_2$O$_6$Fe$_2$P$_2$ (42622) and chalcogenide Fe$X$ representing 11 ($X$ = chalcogenide). FeSe belongs to the 11-family with the simplest crystal structure and the superconductivity can be induced either by charge carrier doping (hole or electron type) or by the applied external pressure. At the ambient pressure, the superconductivity has been observed at $T_c$ = ~8 K for FeSe$_{1-x}$ with $x$ = 0.12 [9]. Hence, the superconductivity in the FeSe system can be induced either by Se deficiency, i.e., FeSe$_{1-x}$ [9], [10], [11] or by Fe excess, i.e., Fe$_{1+x}$Se [12], [13], [14]. Furthermore, the $T_c$ may be greatly enhanced up to 37 K at an applied pressure of about 7 GPa [10]. Generally, the superconductivity in FBS appears after the suppression of magnetic order by a suitable doping [15], [16]. Numerous superconductors have been derived from FeSe in pursuit of finding new superconductors. The superconducting transition is enhanced in the system including $A_x$Fe$_{2-y}$Se$_2$ ($A$ = K, Rb, Tl, etc.) [17], [18] and other organic intercalated superconductors, (Li,Fe)OHFeSe [19], heavily electron-doped FeSe through potassium deposition, specifically, single-layer FeSe/SrTiO$_3$ films with a record high-$T_c$ of ~100 K [20]. The effects of different doping at Se or Fe-sites in the FeSe system have been investigated to comprehend the superconductivity mechanism, including Cu [21], [22] Ni [23], Cr [24], Co [25] at Fe sites, as well as S [25], [26], [27] and Te [25] at Se sites [28]. Further research has shown that substituting Te at Se sites results the highest $T_c$ of up to 14.8 K with an optimal Te content of 50% under the ambient pressure [29]. However, the preparation of single-phase superconducting samples for this 11-family is challenging due to its intricate phase diagram of 11-family, which has numerous stable crystalline forms, such as tetragonal-Fe$_x$Se, hexagonal-Fe$_x$Se, orthorhombic FeSe$_2$, tetragonal-Fe$_x$Se, monoclinic Fe$_3$Se$_4$, and hexagonal Fe$_7$Se$_8$, in which the tetragonal phase generally exhibits the superconductivity with $T_c$ = 8 K [9]. Many reports have revealed that the tetragonal phase coexists with the hexagonal phase, signifying a continual competition between these phases during the growth process [30].

If the growth conditions are not suitable, the hexagonal phases, particularly hexagonal-$Fe_xSe$ and hexagonal $Fe_7Se_8$, emerge rapidly alongside the main tetragonal 11-phase, which adversely affects the superconducting properties [31], [32]. Numerous studies examine the impact of charge carriers, whether hole or electron-doping, on the characteristics of the FeSe system. Large dopant size (Sb) and the equivalent size (Si) dopants at Se sites have been reported in the $FeSe_{1-x}$ system [33], [34], [35] where the hexagonal phase was observed and the onset transition temperature (~9 K) remained nearly constant [33] throughout 5-20% doping contents. These kinds of doping prompted us to investigate the properties of FeSe system with novel dopants. Consequently, we have selected As-doping at Se sites in FeSe system, as there are no existing reports in the literature. Since the 11 family exhibits the same tetragonal structure as the pnictide (FeAs) superconductors, including the 1111 and 122 families. In the FeAs-based compounds, the Fe-atoms establish a square lattice, with arsenic positioned above and below the centre of the Fe-ions square. The substitution of Arsenic with alternative elements is crucial for determining the electronic and magnetic characteristics of the iron-based superconductors, where the hole or electron doping either causes superconductivity or inhibits the magnetic order. To explore the properties of FeSe system through a hole-carrier doping at the Se sites, we have selected this novel arsenic dopant, which is larger than the Se-ionic radius.

In this report, we investigate the effects of As-doping at Se sites in the FeSe(11) material. A series of $FeSe_{1-x}As_x$ bulks is prepared using the conventional synthesis method and characterized through various measurements. Structural and microstructural investigations suggest that a small amount of As-doping is possible at Se sites in the FeSe phase, whereas higher doping contents appear in the form of FeAs as an impurity phase. The resistivity and magnetization measurements confirm the appearance of the onset superconducting transition. Our study depicts that the large size of dopant (i.e. Arsenic) does not support the formation of a tetragonal 11 phase but exhibits the superconducting transition up to 11.6 K.

## 2. Experimental Details

A series of $FeSe_{1-x}As_x$ polycrystalline samples were prepared using conventional two-step solid-state reaction methods with $x = 0.005, 0.01, 0.02, 0.05, 0.1$, and $0.2$. The initial precursors, Fe powder (99.99% purity, Alfa Aesar), Se powder (99.99% purity, Alfa Aesar), and Arsenic (As) chunks (99.99% purity, abcr), were mixed in an agate mortar for 15–20 min according to their nominal compositions. All the chemical processes were performed in a highly pure inert gas atmosphere glove box, exhibiting minimal ppm levels of oxygen and moisture. We prepared 1 g of samples in a single batch. The powders were mixed rigorously and, subsequently, cold-pressed into discs with an 11 mm diameter at a uniaxial pressure of 200 bars. The prepared pellets were sealed into an evacuated quartz tube. The evacuated quartz tube helps to minimize the oxygen and moisture atmosphere surrounding the precursors. In the first step, the evacuated quartz tubes were heated at 600 °C for 11 h in the box furnace. After the completion of the reaction, this quartz tube was opened inside the glove box, and the pellets were reground. The powders were pelletized again, having the same diameter as the first step. Afterwards, the pellets were sealed again in an evacuated quartz tube. In the subsequent step, the sealed tube was heated again at 600 °C for 4 hours within the same furnace. The final disks were blackish and had the same diameter and thickness as the original ones.

The prepared samples, in the form of pellets, were cut into a rectangular bar shape inside the glove box to prevent oxidation, moisture, or contamination. For the phase and structural analysis, XRD measurements have been performed using a Philips X'Pert Pro diffractometer with filtered Cu–Kα radiation (wavelength: 1.5418 Å, power: 30 mA, 40 kV) and a PIXcel1D position scintillation detector. A Zeiss Ultra Plus field-emission scanning electron microscope equipped with the EDS microanalysis system by Bruker mod is employed for the detailed microstructural analysis and the elemental mapping of the constituent elements. The transport properties, particularly the resistivity, were measured across a temperature range of 7–300 K in the absence of a magnetic field, employing a four-probe approach with a closed-cycle refrigerator (CCR). The data were collected using a low-warming procedure to study the temperature dependence of resistivity at current I = 20 mA. Magnetic measurements were performed utilizing a VSM connected to a Quantum Design PPMS, applying a field of 50 Oe in zero-field cool (ZFC) and field-cool (FC) modes across a temperature range of 5-30 K.

## 3. Results and discussion

All the prepared samples are characterized using X-ray powder diffraction, as illustrated in Figure 1. Due to the utilization of a Pt sample holder, these patterns also encompass the Pt peaks. The main peak 101 in Figure 1 corresponds to the tetragonal phase of FeSe, referring to the superconducting phase. Low As-doped samples ($x$ = 0.005 and 0.01) have the main peak of the tetragonal phase with space group P4/nmm, while a significant presence of the hexagonal ($Fe_7Se_8$) phase is also detected. With further increase of As-doping contents ($x$ = 0.02-0.1), the hexagonal phase is dominated over the tetragonal phase, becoming the primary phase seen. Interestingly, for $x$ = 0.2, the tetragonal phase re-emerges as the predominant phase, while the concentration of the hexagonal phase is reduced compared to other intermediate doping levels. However, at this doping level, the concentration of the FeAs phase commences to increase. The presence of substantial quantities of hexagonal $Fe_7Se_8$ and FeAs phases could not be favourable for the superconducting properties. This suggests that the larger atomic size of the dopant is not supportive of the tetragonal phase formation of FeSe. Along with these impurities, additional peaks are detected from the Pt sample holder used for our XRD instrument. The calculated lattice parameters are $a$ = 3.764 Å, $c$ = 5.521 Å for $x$ = 0.01; $a$ = 3.763 Å, $c$ = 5.529 Å for $x$ = 0.05; $a$ = 3.763 Å, $c$ = 5.525 Å for $x$ = 0.2. The obtained lattice parameters for FeSe are $a$ = 3.765 Å, and $c$ = 5.518 Å, as reported elsewhere [36]. Hence, the lattice parameter '$c$' is increased for the low amount of As-contents ($x$ ≤0.05), which confirms the successful insertion of As at Se sites. One can note that the lattice '$c$' of high As-doping contents (i.e., $x$ = 0.2) is nearly identical to that of the samples $x$ = 0.01-0.05. It means the arsenic may no longer be effectively incorporated into the FeSe lattice beyond a certain doping level, namely, $x$ = 0.05, and manifest as the FeAs phase. One can note that the presence of several phases may result in substantial error bars for these obtained lattice parameters.

To examine the distribution of the constituent elements inside these samples, the bulks were polished using sandpaper from low grade to extremely fine grade without the application of any lubricant. The elemental mapping for various bulk samples is collected for all samples and illustrated in Figure 2(a)-(o) for the selected samples $x$ = 0.005, 0.05, and 0.2, respectively. The sample $x$ = 0.005 exhibits a nearly homogeneous distribution of the constituent elements Fe, Se, and As; however at many places, the inhomogeneities are also evident, confirming the existence of $Fe_7Se_8$ (hexagonal phase) and FeAs phase (Figures 2(a)-(e)). The inhomogeneity of the constituent elements was more pronounced for the sample $x$ = 0.05, suggesting a significant presence of impurity phases (Figures 2(f)-(j)). As the concentration of As-dopants increases, the sample exhibits greater inhomogeneity, as depicted in Figures 2(k)-(o) for $x$ = 0.2. Similar elemental mappings are also observed for other samples. These results confirm the presence of hexagonal and FeAs phases alongside the tetragonal phase, and also suggest that As-doping increases the inhomogeneity of the constituent elements. The observations were consistent with the above-discussed XRD data.

The polished samples are also used for the microstructural analysis and the backscattered scanning electron (BSE) images are collected on various magnification scales for all these samples to reveal chemical contrast. Figure 3(a)-(i) depicts BSE images of the selected samples $x$ = 0.005, 0.05 and 0.2 arranged from low to high-resolution scale. These images exhibit light grey, white, and black contrasts that correspond to $FeSe_{1-x}As_x$, hexagonal phase $Fe_7Se_8$, and pores respectively. In some places, the black contrast refers to the FeAs phase. Figure 3 suggests that the samples are not very compact and many pores do exist. It appears that many grains are not well connected due to the presence of micro- or nanopores, and impurity phases such as FeAs or $Fe_7Se_8$ between them. The observed images of all these samples confirm the presence of hexagonal and FeAs phases, i.e., inhomogeneous microstructure, supporting the analysis of the elemental mapping.

The temperature dependence of the normal state resistivity is illustrated in Figure 4(a) for all the prepared bulks within the temperature range of 7–300 K. The resistivity behaviour is almost the same as reported for $FeSe_{1-x}$ [9], [37], however, a systematic variation of the resistivity corresponding to different As-doping contents is not evident, which could be due to the formation of a hexagonal phase promoted by As-dopants. The resistivity is increased for the samples $x = 0.01$ and 0.02 in comparison to $x = 0.005$, however it is decreased for $x = 0.05$. The sample $x = 0.1$ has exhibited a significantly high resistivity value across the whole temperature range which could be due to the main phase of the hexagonal phase for this sample, as shown in Figure 1. With further increased level of As-doping, specifically $x = 0.2$, the resistivity is lower than that of the sample $x = 0.1$, although higher than that of other samples. This non-systematic behaviour may result from the existence of several impurity phases and the competition between hexagonal and tetragonal phases, as corroborated by the structural and microstructural studies.

To understand the superconducting transition of these samples, the low-temperature resistivity behaviour is shown in Figure 4(b) within the temperature range 7-20 K. Interestingly, all samples exhibit a drop in the resistivity around 11 K, nevertheless, no zero resistivity is observed until the measured temperature up to 7 K. A slight increase of the onset transition is observed for low As-doped FeSe, specifically for $x = 0.005$, 0.01, and 0.02. At high As-doping levels, the onset transition temperature is slightly reduced. Interestingly, the zero resistivity is not reached, which could be attributable to the existence of non-superconducting phases. The sample $x = 0.20$ reaches nearly zero but doesn't exhibit complete zero resistivity up to 8 K. The maximum value of critical temperature is observed for a 2% doping level ($x = 0.02$) with $T_c^{onset}$ =11.5 K. The basis for the definition of the onset critical temperature is the deviation from the linear behaviour of the normal state resistivity ($\rho$).

To confirm the superconducting behaviour, we have measured the magnetic moment of two specific samples: $x = 0.05$ and $0.2$ within the temperature range of 5-30 K at a magnetic field of 50 Oe, as depicted in Figure 5. We have presented a normalized magnetic moment to facilitate a comparative study of these samples. The existence of the impurity phases results a significant background in the magnetic moment measurements, as shown in Figure 5, which agrees well with the above-discussed analysis. Both samples, $x = 0.2$ and $0.05$, exhibit the same superconducting transition at approximately 8.5 K, suggesting that the actual As-contents in these samples may be nearly identical. The structural analysis agrees well with this observation. The obtained $T_c$ from these measurements is around 2-3 K lower than that of the resistivity measurements. It could be possible due to inhomogeneity and the existence of the non-superconducting phase in As-doped samples.

To summarize our findings, we have plotted $T_c^{onset}$ value from the resistivity measurements and the room temperature resistivity ($\rho_{300K}$) with As-doping contents ($x$) in Figure 6. A slight variation in the room temperature resistivity is observed for low As-doping contents, specifically $x = 0.005, 0.01, 0.02$ and $0.05$, but it increases significantly for high As-doping contents, as depicted in Figure 6(a). The $T_c^{onset}$ is slightly improved for low As-doping contents from 10.8 K to 11.5 K. Further increases in As-doping contents, i.e., 5-10%, slightly reduce the transition temperature, however a significant reduction is observed for the $x = 0.2$ sample, as illustrated in Figure 6(b). These analyses suggest that As-doping works effectively for low doping contents up to 5%, but high amounts of doping reduce the transition temperature and enhance the inhomogeneity of the FeSe$_{1-x}$As$_x$ bulks. More studies involving a large amount of As-doping ($x>0.2$) and the single crystal growth of FeSe$_{1-x}$As$_x$ could be helpful for a clear understanding of the influence of As-doping on the properties of FeSe materials.

## 4. Conclusion:

We have prepared a series of FeSe$_{1-x}$As$_x$ ($x = 0.005, 0.01, 0.02, 0.05, 0.1$ and $0.2$) bulks and investigated the arsenic (As) doping effect on the properties of the FeSe system. Structural and microstructural analysis confirms the inhomogeneity of the samples attributable to the presence of the impurity phases. Arsenic doping contents induce superconductivity with the highest $T_c^{onset}$ = 11.5 K for 2% As-doping, but zero resistivity is not observed within the measured temperature range. However, at higher As-doping concentrations, the superconducting transition is reduced. Our study suggests that a low amount of As-doping (up to 5%) can be effective for FeSe to induce the superconducting properties. Further comprehensive research is required to understand the influence of arsenic dopants on the FeSe structure and its effect on superconducting characteristics.

## Acknowledgments:

The work was funded by SONATA-BIS 11 project (Registration number: 2021/42/E/ST5/00262) sponsored by National Science Centre (NCN), Poland. SJS acknowledges financial support from National Science Centre (NCN), Poland through research Project number: 2021/42/E/ST5/00262.

**Figure 1:** Powder X-ray diffraction patterns (XRD) at room temperature for the prepared samples FeSe$_{1-x}$As$_x$ at ambient pressure with $x$ = 0.01, 0.05, 0.1, and 0.2. FeAs, hexagonal (Fe$_7$Se$_8$) and Pt phases are observed as the impurity phases. The Pt impurity phase is observed from the XRD sample holder.

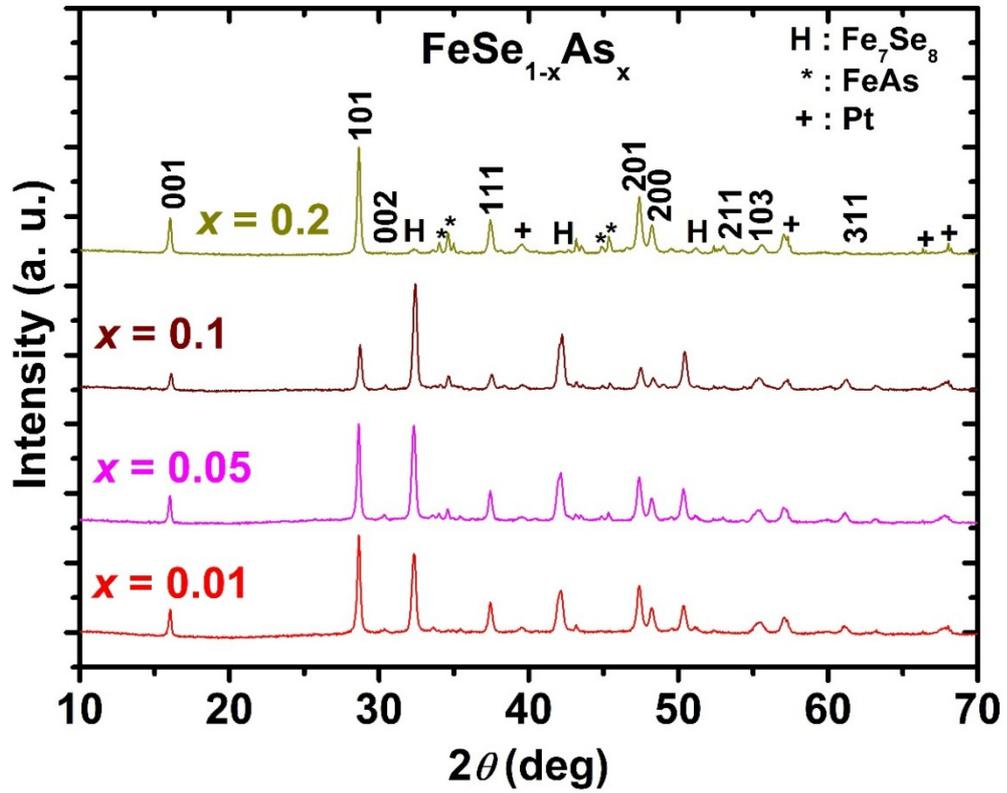

**Figure 2:** Elemental mapping for the constituent elements of FeSe$_{1-x}$As$_x$ polycrystalline samples: **(a)-(e)** $x = 0.005$; **(f)-(j)** $x = 0.05$; and **(k)-(o)** $x = 0.2$. For each sample, the first image depicts a Scanning Electron Microscope (SEM) image, and the last image represents the combined image of Fe, Se and As-elements. The rest of the images are the elemental mapping of the individual elements.

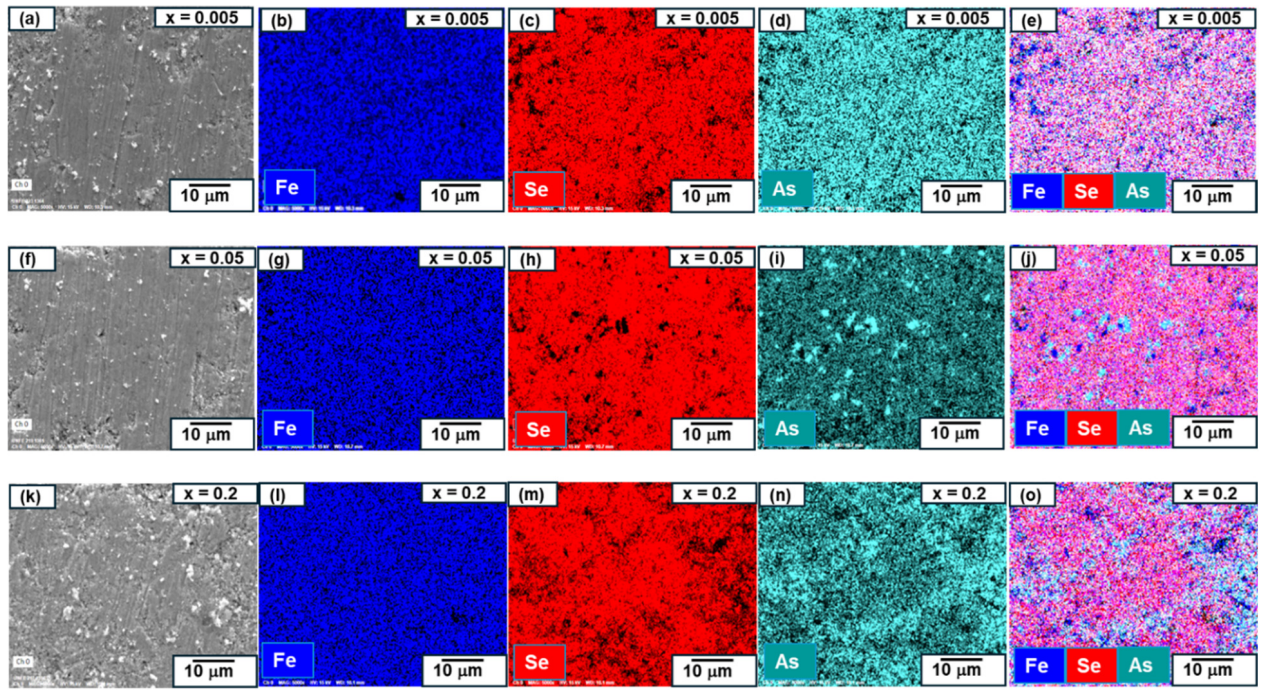

**Figure 3:** Backscattered electron image (BSE; AsB) of the samples with **(a–c)** $x = 0.005$, **(d–f)** $x = 0.05$ and **(g–i)** $x = 0.2$ samples. Bright, light grey, and black contrasts correspond to $Fe_7Se_8$ (hexagonal phase), $FeSe_{1-x}As_x$, and pores, respectively. The black contrast may indicate the presence of the FeAs phase in certain locations.

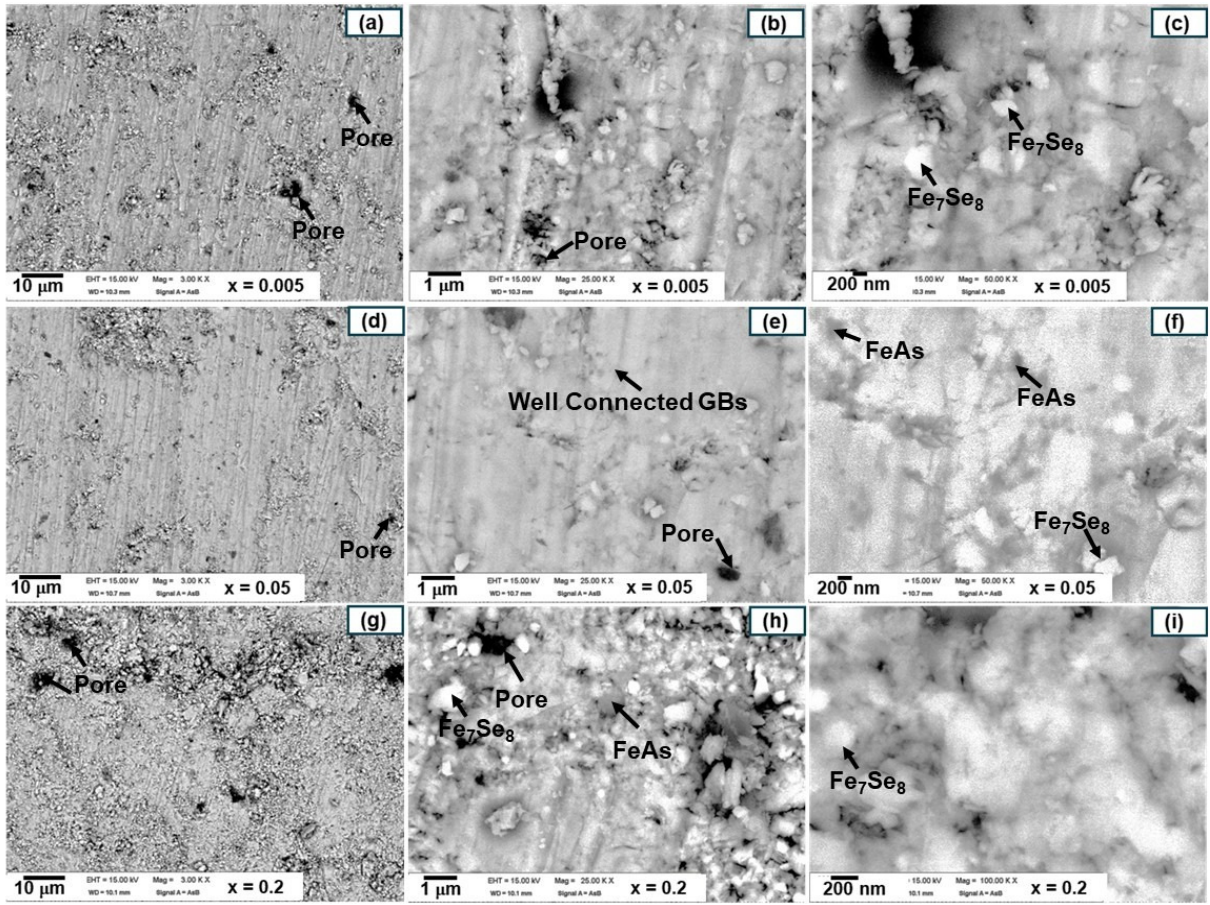

**Figure 4: (a)** The variation of resistivity with temperature for all the samples up to the room temperature (300 K). **(b)** The variation of low-temperature resistivity with temperature ranges from 7 to 20 K. The insets of these figures (a) and (b) illustrate the resistivity variation for the doping content $x = 0.1$.

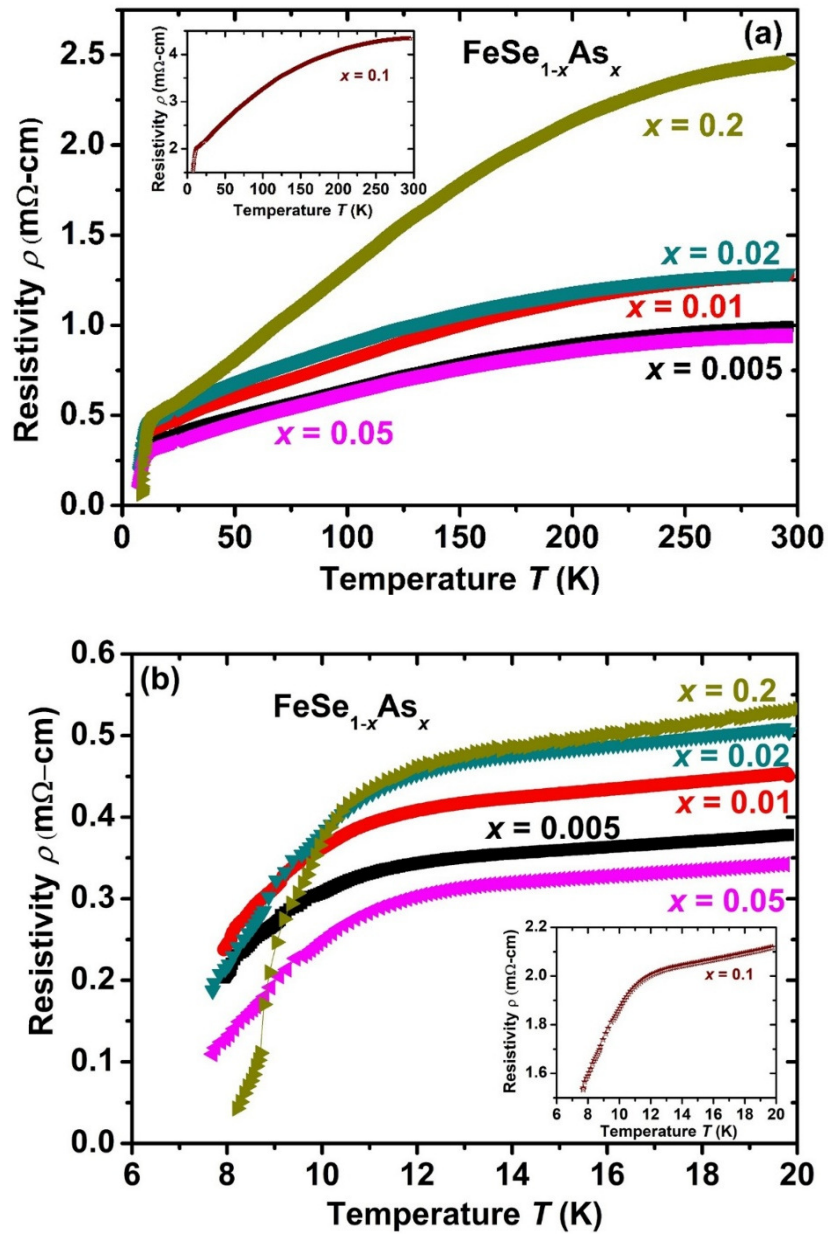

**Figure 5:** The temperature dependence of the normalized magnetic moment ($M/M_{5K}$) with an applied magnetic field of 50 Oe for FeSe$_{1-x}$As$_x$ with $x = 0.05$ and $x = 0.2$ in zero-field cooling (ZFC) and field cooling (FC) modes. The inset figure shows the variation of the normalized magnetic moment for the sample $x = 0.05$.

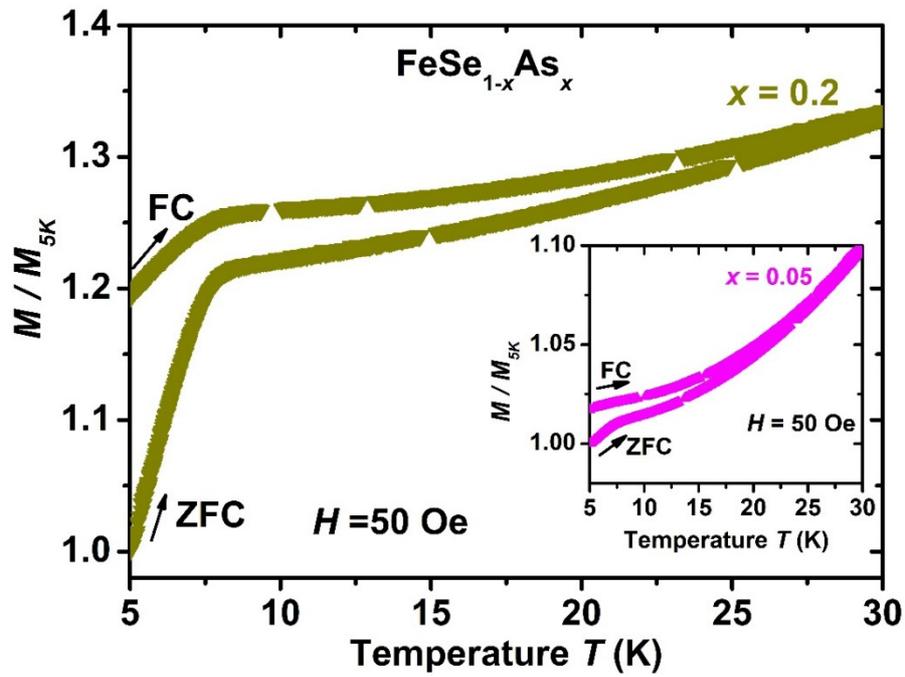

**Figure 6:** The variation of **(a)** room temperature resistivity ($\rho_{300K}$) and **(b)** the onset critical temperature $T_c^{onset}$ from the resistivity measurements with As-doping content ($x$).

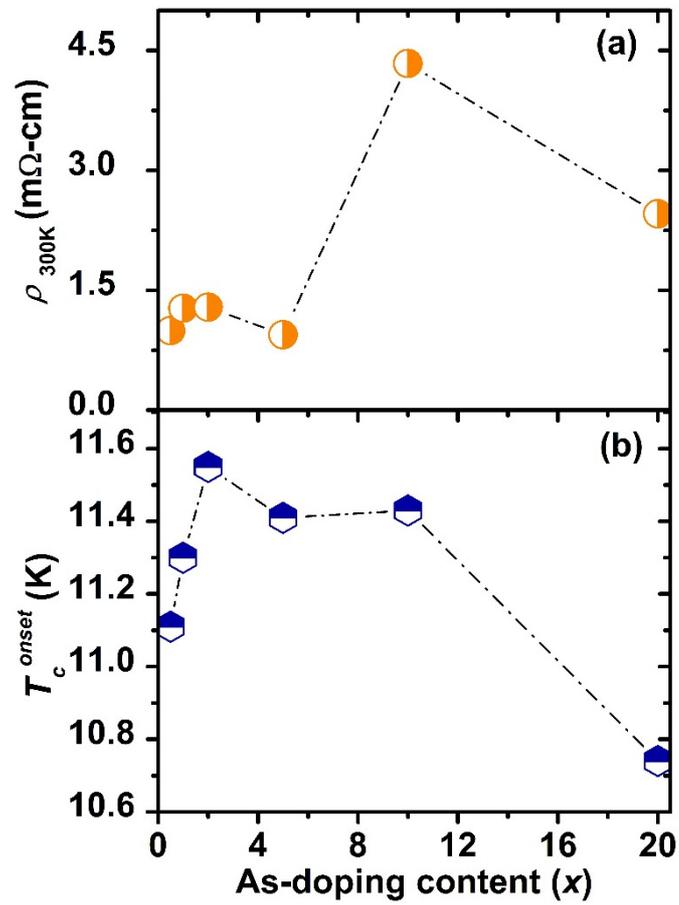